# Charge transport layer dependent electronic band bending in perovskite solar cells and its correlation to device degradation

Junseop Byeon[1,2,†], Jutae Kim[1,†], Ji-Young Kim[3], Gunhee Lee[1], Kijoon Bang[1], Namyoung Ahn*[1], and Mansoo Choi*[1,2]

[1]Global Frontier Center for Multiscale Energy Systems, Seoul National University, Seoul 08826, Republic of Korea.
[2]Department of Mechanical and Aerospace Engineering, Seoul National University, Seoul 08826, Republic of Korea.
[3]Advanced Analysis Center, Korea Institute of Science and Technology (KIST), Hwarangno 14-gil 5, Seongbuk-gu, Seoul 02792, Republic of Korea.

* To whom correspondence should be addressed.
E-mail: Mansoo Choi (mchoi@snu.ac.kr), Namyoung Ahn(nyny92@snu.ac.kr)

† These authors contributed equally to this work.

# Abstract

Perovskite solar cells (PSCs) have shown remarkably improved power-conversion efficiency of around 25%. However, their working principle remains arguable and the stability issue has not been solved yet. In this report, we revealed that the working mechanism of PSCs is explained by a dominant p-n junction occurring at the different interface depending on electron transport layer, and charges are accumulated at the corresponding dominant junction initiating device degradation. Locations of a dominant p–n junction, the electric field, and carrier-density distribution with respect to electron-transport layers in the PCS devices were investigated by using the electron-beam-induced current measurement and Kelvin probe force microscopy. The amount of accumulated charges in the devices was analyzed using the charge-extraction method and the degradation process of devices was confirmed by SEM measurements. From these observations, we identified that the dominant p-n junction appears at the interface where the degree of band bending is higher compared to the other interface, and charges are accumulated at the corresponding junction where the device degradation is initiated, which suggests that there exists a strong correlation between PSC working principle and device degradation. We highlight that an ideal p–i–n PSC that can minimize the degree of band bending should be designed for ensuring long-term stability, via using proper selective contacts

## Introduction

Perovskite solar cells (PSCs) have attracted worldwide attention in recent years because of their soaring performance. The power-conversion efficiency (PCE) of the PSC exceeded that of its direct predecessors (dye-sensitized solar cells and organic photovoltaics).[1,2] Many researchers have investigated the working principle of a PSC and explained the carrier dynamics in a PSC based on semiconductor physics, in which the perovskite layer could be considered as an intrinsic layer.[3, 4] In other words, the entire perovskite layer could be assumed to be a space-charge region that can separate electron–hole pairs and transfer the carriers to charge-selective layers, as observed in amorphous-silicon solar cells. However, several other studies reported that perovskite layers behaved as different types of semiconductor (p, i, or n) depending on their ionic composition[5-8] and the types of adjacent charge-selective layers.[9] These studies indicate that the exact working mechanism of PSCs has not been settled yet.

The stability issue of PSCs has not been solved as well, although numerous studies were done on degradation phenomena.[10-20] Oxygen and water molecule were pointed out as extrinsic origins of degradation of PSCs[10-13]. Also, charges such as photo-carriers and ions have been identified as intrinsic sources of degradation.[14-18] Especially, it was suggested that the irreversible degradation could be caused by trapped charges in the perovskite film combined with $O_2$ or $H_2O$ (or both $O_2$ and $H_2O$) molecules.[14] Although extrinsic origins can be prevented to some extent by additional encapsulation[19,20], there is no way to completely block them for ensuring a long-term stability compatible to commercialized solar cells and the encapsulation itself could not prevent the trapping of charges that would be an intrinsic source of device degradation. As such charge behaviors including charge dynamics and trapping should be relied on the working principle, it is important not only to elucidate the precise working mechanism of PSCs, but also identify a clear relationship between working mechanism and device degradation.

Herein, we verified that solar cells based on methylammonium lead iodide ($MAPbI_3$) could be represented not by a p–i–n junction but a p–n junction dominating at a specific interface depending on the kind of ETLs. This ETL dependent p–n junction type working principle is deeply related with charge accumulation behavior in the PSCs that affects degradation process. To dig into ETL dependent working principle of PSCs, the devices are investigated by analyzing electron-beam-induced current (EBIC) and Kelvin probe force microscopy (KPFM) measurements. In addition, to investigate the principle of device degradation, we measured the accumulated charges (their amount and type (hole or electron)) when different types of ETLs were employed and analyzed scanning electron microscope (SEM) images of degraded devices that aged under light irradiation in ambient conditions. The results indicate that the dominant p–n junction occurs at the different interface depending on which kind of ETL is used and the charges mainly accumulate at the dominant p-n junction interface where the device degradation is initiated. A polarity of accumulated charges is also dependent on the location of the dominant p-n junction, for example, holes are accumulated near ETL layer when the p-n junction occurs near ETL and electrons are accumulated near HTL layer when p-n junction occurs near HTL. We discovered, for the first time, that the performance deterioration of the PCS was highly correlated with the dominant p–n junction forming depending on charge transport layers and the amount of accumulated charges at the junction. From all these observation and analysis, we propose that it is necessary to minimize band bending at the interfaces by designing an ideal p–i–n type of PSC for long-term stability via using proper selective contact.

## Results and Discussion

A band diagram associated with the electronic band structure of a semiconducting material in a solar cell is a very useful visual aid for understanding the working principle and junctions of the device. Cross-sectional EBIC measurements can be used to investigate the junction structures among the layers of the solar cell,[21-23] and the obtained current signals provide information on the location of the built-in electric field (E-field), defects, and diffusion length of photo-carriers.[24-26]

We conducted cross-sectional EBIC measurements for MAPbI$_3$ PSCs fabricated with three different ETLs: a titanium dioxide (TiO$_2$) layer, a fullerene (C$_{60}$) layer, and a TiO$_2$/C$_{60}$ bilayer. The other layers, such as the perovskite (MAPbI$_3$) layer and the hole-transport layer (HTL; 2,2′,7,7′-tetrakis(*N*,*N*′-di-*p*-methoxyphenylamine)-9,9'-spirobifluorene, or spiro-MeOTAD), were the same for all three devices. The architecture of the devices can be summarized by the layered structure of FTO/ETL/MAPbI$_3$/HTL/Au (where FTO stands for fluorine-doped tin oxide). During the measurements, an electron-beam (E-beam) with high energy was focused on the cross-sectional plane of the sample to generate excitons. The generated charges flowed through the external short circuit, and the amount of charge flow was measured at the same time, as illustrated in **Fig. 1a**. The generated current values from the E-beam and their corresponding positions can be visualized by overlapping the cross-sectional SEM image and the measured current signals. **Fig. 1b–d** show the EBIC results laid over corresponding cross-sectional SEM images of devices with different ETLs. The bright green color inside a MAPbI$_3$ layer indicates higher EBIC signals. For the TiO$_2$-based device, shown in **Fig. 1b**, the bright green region was formed mostly at the bottom of the MAPbI$_3$ layer, near the interface with the TiO$_2$ layer. The observed high EBIC signals at the interface between the perovskite light-absorbing layer and the TiO$_2$ ETL are in agreement with previously reported results.[21,27] On the other hand, for the C$_{60}$-based and TiO$_2$/C$_{60}$-based devices, the green region was mainly located near the spiro-MeOTAD layer, as shown in **Fig. 1c** and **1d**. It is important to note that the current signals and the brightness of the green color in this study indicate the relative current, not the absolute values of the current.

As the EBIC signals can identify the location of the dominant junctions in the device[23], line profiles of EBIC signals perpendicular to the layers were obtained to analyze the junctions associated with the employed ETL layer(see **Fig. 1e–g).** It was confirmed in the TiO2 ETL based device that, unlike the weak EBIC signals adjacent to the MAPbI$_3$/spiro-MeOTAD interface, the EBIC signals were extremely biased toward the TiO$_2$/MAPbI$_3$ interface (**Fig. 1e**). This means that a substantial amount of the generated excitons near the TiO$_2$ interface dissociated and transferred to the electron-transport layer. In other words, there was a strong built-in E-field at the dominant junction (depletion region) near the interface between TiO$_2$ and MAPbI$_3$ layers. On the contrary, for the TiO$_2$/C$_{60}$-based and C$_{60}$-based devices, EBIC signals of relatively high intensity appeared in the MAPbI$_3$ layer near the spiro-MeOTAD interface. These results indicate that the dominant junction (depletion region) now forms at the spiro-MeOTAD interface when the MAPbI$_3$ layer is sandwiched between C$_{60}$ and spiro-MeOTAD layer (**Fig. 1f and 1g**, respectively), clearly revealing that the location of the dominant junction changes depending on the ETL material. For the TiO$_2$/C$_{60}$-based device, high EBIC signals at some region were obtained near the ETL interface, like for the TiO$_2$-based device, owing to a local pinhole in the C$_{60}$ layer or partial peeling of the C$_{60}$ layer (**Fig. S1**). The EBIC signals of the C$_{60}$-based device were distributed to be relatively uniform throughout the entire MAPbI$_3$ layer when compared to signals of the other devices (**Fig. 1g**). This means a longer carrier diffusion length in this case, implying a thicker depletion region (This will be discussed later).

Through the EBIC measurements, we confirmed that specific locations of the depletion region in the MAPbI$_3$ PSCs depended on the material of the neighboring ETL. In order to understand the more-detailed solar-cell physics underlying the observed behavior, the electrical properties across the entire device must be determined. To estimate the E-field inside the MAPbI$_3$ PSCs, the local contact potential differences (CPDs) in the devices with different ETLs (TiO$_2$ and C$_{60}$) were measured by cross-sectional KPFM, as illustrated in **Fig. 2a**. The surfaces of the devices were treated by focused ion beam (FIB) milling to obtain smooth cross-sectional planes, which reduced signal errors in the surface potential that could be caused by the rough topography.[28-30] In order to analyze the electrical properties, the KPFM measurements of the devices were conducted under LED illumination in both the open-circuit (OC) state and short-circuit (SC) state. **Fig. 2b** and **2e** show the averaged CPD values obtained by scanning across the layers at OC (red line) and SC (blue line) states for the TiO$_2$-based and C$_{60}$-based devices, respectively. Both topographic (top) and CPD images (middle and bottom) of the two devices in the SC and OC states are shown in **Fig. S2**. Their topographic images show low surface roughness implying no artifact of CPD values resulting from local surface morphologies.[28-30]

When a semiconducting material with a higher Fermi level (relatively n-type) is placed adjacent to another semiconducting material with a lower Fermi level (relatively p-type), the vacuum level (VL) is bent to align two Fermi levels. Such VL shift forms an electrical built-in potential ($V_{BI}$), which can be directly identified from CPD values measured in the SC state in both dark and illuminated conditions.[30,31] As shown in **Fig. 2b** and **2e**, the CPD for the SC state decreased from the value of the ETL toward the value of the HTL, on the other hand, for the OC state, it gradually increased from the ETL value toward the HTL value owing to the separate quasi-Fermi levels caused by electron and hole accumulation in both $TiO_2$ and $C_{60}$-based devices. For the $TiO_2$-based device (**Fig. 2b**), the CPD measured under illumination in the SC state (blue line) decreased mainly at the $TiO_2$/$MAPbI_3$ interface. The CPD profile in the OC state showed a steep slope at the $TiO_2$/$MAPbI_3$ interface, while the spiro-MeOTAD side showed a slight increase (**Fig. 2b**), providing a clear evidence of the formation of a dominant p–n junction at the $TiO_2$ interface. Contrary to the case of $TiO_2$-based device, the $C_{60}$-based device showed a larger VL shift at the spiro-MeOTAD side in both OC and SC states as compared to the $C_{60}$ interface (**Fig. 2e**), which is indicative of the dominant p–n junction occurring at the spiro-MeOTAD interface. It is also noted that the CPD profile in the OC state exhibited a relatively uniform rise in the $MAPbI_3$ layer without steep slope, which could influence less charge accumulation than that for the $TiO_2$-based device. (discussed later) These results are in good agreement with in EBIC measurements (**Fig. 1**). Additionally, the difference between CPD values in the OC and SC states at the gold electrode correspond to an actual open-circuit voltage,[28,30] and it was ~670 mV and ~740 mV in the $TiO_2$-based and $C_{60}$-based devices, respectively. Both devices had similar $V_{OC}$ values during the PCE measurements, as shown in **Fig. S3**. The values from the KPFM measurements, however, were smaller than the values measured under 1 sun illumination because of the different light sources and intensities (**Fig. S3**).

To determine the relative magnitude of the E-field across the device, we calculated the normalized E-field from the CPD distribution measured in the OC state under illumination conditions based on the following equation (**Fig. 2c** and **2f**):[28-30]

$$E(x) = -\frac{d}{dx}\text{CPD}(x) \qquad (1)$$

For the TiO$_2$-based device, there was a strong E-field at the TiO$_2$ interface, while a weak E-field was uniformly distributed in the MAPbI$_3$ layer and at the MAPbI$_3$/spiro-MeOTAD interface, as shown in **Fig. 2c**. The field distribution indicates that charges mostly accumulated at the TiO$_2$ interface, forming an open-circuit voltage. On the other hand, for the C$_{60}$-based device, a strong E-field appeared at the MAPbI$_3$/spiro-MeOTAD interface rather than at the C$_{60}$/MAPbI$_3$ interface, as shown in **Fig. 2f**. This again confirms that the location forming dominant p-n junction depends on the choice of ETL. It is also emphasized that the profile of normalized electric field for C$_{60}$-based device is more uniform than the case for TiO$_2$-based device, which is found to be correlated with the difference in degradation speed of two devices (discussed later).

We also estimated the carrier-density distribution inside the devices from the KPFM results by using the following Poisson's equation: [29,30]

$$\rho(x) = \varepsilon_0 \varepsilon_r \frac{d}{dx} E(x) = -\varepsilon_0 \varepsilon_r \frac{d^2}{dx^2} \Delta CPD(x) \;, \tag{2}$$

where $\rho(x)$ is the carrier density, $E(x)$ is the E-field, $\Delta CPD$ is the difference between CPD values measured at the SC and OC states, $\varepsilon_0$ is the dielectric constant of vacuum, and $\varepsilon_r$ is the dielectric constant of the perovskite (MAPbI$_3$) lay10er.

The carrier density profiles were obtained from equation (2) for the TiO$_2$-based (**Fig. 2d**) and C$_{60}$-based (**Fig. 2g**) devices, respectively. For the TiO$_2$-based device, holes were intensively accumulated at the TiO$_2$ interface, and the net charge of the MAPbI$_3$ layer was also positive owing to this accumulation (**Fig. 2d**). The charge-carrier profile of the TiO$_2$-based device exhibited a similar trend to those of previously reported results,[30] but the profile of the C$_{60}$-based device presented accumulated electrons mainly at the spiro-MeOTAD interface rather than at the C$_{60}$ interface. The net charge of the entire MAPbI$_3$ layer in the C$_{60}$-based device was negative in contrast to the case of TiO$_2$ based device and its absolute value was lower than that of the TiO$_2$-based device. These charge distribution is the result of electric field distributions shown in Fig.2c and 2f and closely correlated with the behavior of device degradation (discussed later).

We investigated the location of the dominant junction, VL shift, E-field under illumination, and carrier density distribution from the EBIC and KPFM results. From these, we confirmed that these device properties depended on the neighboring charge-transport layers as the actual devices functioned rather like p–n solar cells at the dominant interface, not an ideal p–i–n. It is desirable to determine a detailed model to explain the charge transport layer dependent working principle of PSCs that we found. We propose a model based on our experimental results to explain the PSC junction configuration and related physics, in which band bending between two adjacent semiconducting materials is the key factor.

**Fig. 3** presents band diagrams of two different $TiO_2$-based and $C_{60}$-based devices in both SC and OC states; the band bending required to align the different Fermi levels of the two semiconducting materials were taken into consideration. Fermi level of n-type ETL is close to the conduction band edge. The higher energy level of the conduction band (~ Fermi level) of the $TiO_2$ ETL than that of the $C_{60}$ ETL resulted in larger band bending in the $TiO_2$-based device.[32,33] In the SC state, the built-in potential caused by band bending played a decisive role in the separation and collection of photo-generated charge carriers. As a result, the $TiO_2$/$MAPbI_3$ interface was the dominant location of the carrier drift in the $TiO_2$-based device (**Fig. 3a**). In the $C_{60}$-based device, carrier drift occurred mainly at the $MAPbI_3$/HTL interface (**Fig. 3b**). The OC state resulted in holes accumulating at the ETL/$MAPbI_3$ interface of the $TiO_2$-based device (**Fig. 3c**) and electrons accumulating at the $MAPbI_3$/HTL interface of the $C_{60}$-based device (**Fig. 3d**). In particular, many holes accumulated at the $TiO_2$/$MAPbI_3$ interface, while a small number of electrons resided at the $MAPbI_3$/HTL interface in the $TiO_2$-based device. In contrast, in the $C_{60}$-based device, more electrons resided at the $MAPbI_3$/HTL interface than the number of holes accumulating at the $C_{60}$/$MAPbI_3$ interface. Such carrier dynamics are consistent with our observations from the EBIC and KPFM results.

In order to quantify absolute accumulated charges of the operating device under light illumination in the OC state, we directly measured the amount of accumulated charges in the three different devices through the charge-extraction method.[34-36] The accumulated charges in a device in the OC state and under illumination were extracted after a certain delay, switching to the SC state and dark conditions. A shorter delay yielded more extracted charges, and the details are shown in **Fig. S4**. **Fig. 4a** shows the amount of the extracted charges as a function of the delay time up to 4s for the three different devices. The $TiO_2$-based device exhibited the largest amount of extracted charges regardless of the delay time, which is consistent with the KPFM results shown in **Fig. 2d** and **2g**. It is obvious that charge accumulation is a direct result of the PSC working principle mentioned above. The amount of extracted charges for the three different devices is shown as a function of whole delay times in **Fig. S5**. As the charge accumulation has been reported to be closely related to the measured capacitance,[37,38] we investigated the surface capacitance using electrochemical impedance spectroscopy (EIS) (**Fig. S6**).[39] The surface charge capacitance of the $TiO_2$-based device had the highest value when compared to the $C_{60}$-based and $TiO_2/C_{60}$-based devices, which is also consistent with the above results of charge extraction. The amount and location of charge accumulation is not only a direct result of PSC working principle, but strongly influences the device degradation.

To elucidate the relationship between charge accumulation as a result of PSC working principle and device degradation, we measured the time evolution of the normalized PCEs for three un-encapsulated devices under 1 sun illumination and ambient conditions (**Fig. 4b**). The devices were in the OC state and thus had accumulated carriers during the measurements as shown in Fig.2 and Fig.4a. The results showed a close correlation between the accumulated charges and light-induced stability in air; this is the first documented observation of such correlation. As can be seen in **Fig. 4a** and **4b**, the drop in PCE is the smallest for $C_{60}$ based device that showed the smallest extracted charge and the largest for $TiO_2$ based device that had the largest extracted charge. This result is consistent with the mechanism of trapped charge driven degradation.[14-16] We also confirmed the effect of the accumulated charges on the degradation starting location by analyzing the degradation patterns using FIB–SEM measurements. **Fig. 4c-e** show the time evolution of cross-sectional SEM images during the degradation of the $TiO_2$-based, $TiO_2/C_{60}$-based, and $C_{60}$-based devices, respectively. These SEM images show remarkable differences among the three devices with respect to the initial location of degradation of the $MAPbI_3$ layer. In the $TiO_2$-based device, the $MAPbI_3$ layer began to disintegrate mostly from the interface with the $TiO_2$ layer (**Fig. 4c**), while the $MAPbI_3$ layer in the $TiO_2/C_{60}$-based and $C_{60}$-based devices broke down much closer from the interface with the spiro-MeOTAD layer (**Fig. 4d** and **4e**). Interestingly, the degradation locations were identical to the location of the dominant p–n junctions where charges were accumulated (confirmed from the EBIC and KPFM measurements (**Fig. 1** and **2**). These results clearly revealed that the accumulated charges (the amount and location) of photovoltaic devices were deeply associated with device stability as suggested in previous studies.[14-16,40,41] This suggests that there exists a strong correlation between PSC working principle and device stability since the charge accumulation behavior is a direct result of PSC working principle.

Depending on different ETLs, the dominant p-n junction formed at the different interface which had a higher degree of band bending than the other interface and most of charges were accumulated at the corresponding junction where the degradation was initiated. This suggests that the working mechanism of PSCs based on the formation of the dominant p-n junction depending on ETL should be strongly correlated with device stability through accumulated charges that can be said as a mediator playing between the working mechanism and device stability

Based on these results, we propose an ideal device for ensuring both performance and stability, as illustrated in **Fig. 5a** and **5b**. The energy band diagram of an ideal p–i–n PSC is similar to that of an a-Si:H/µc-Si:H solar cell described by the drift model (a-Si:H stands for hydrogenated amorphous silicon; µc-Si:H stands for hydrogenated microcrystalline silicon).[47,48] This ideal p–i–n PSC is expected to exhibit high performance and stability because there are a uniform carrier and ion distribution under all circumstances. Moreover, there would be little accumulation of charges because of the homogeneous potential difference and uniform E-field in the device. When the entire perovskite layer acts as an intrinsic layer (depletion region), we can expect not only faster carrier separation and extraction, but also less degradation owing to the absence of regions with high density of charges. Proper selective contacts should be sought to achieve this.

## Conclusions

This study provided an insight into the junction that can form in MAPbI$_3$-based PSCs and the device degradation caused by accumulated charges for the TiO$_2$-based, TiO$_2$/C$_{60}$-based, and C$_{60}$-based devices. The results of the EBIC measurements demonstrate that the location of the dominant p–n junction depends on the selective contacts. The band alignment, E-field under illumination, and carrier- density distribution in the devices were visualized using the cross-sectional KPFM data. We found that charges mainly accumulated at a specific interface corresponding to the dominant p–n junction (that depended on ETL) and directly measured the amounts of the accumulated charges using the charge extraction method. These investigations allowed us to clearly explain the different locations of initial degradation sites in the MAPbI$_3$ layer, which was evidenced by SEM measurements for the degraded devices. Based on all our observations, we found for the first time that these should be a strong correlation between the charge transport layer dependent working mechanism of PSCs and device degradation. Our study suggests a future direction for ensuring both high performance and stability, which should be a design of p–i–n type PSC that could minimize band bending at the interfaces by using proper selective contact.

**Experimental Section**

*Fabrication of perovskite solar cells*

Patterned fluorine-doped tin oxide (FTO) glass substrates (AMG, Korea) were cleaned and rinsed sequentially using acetone, isopropanol, and distilled water. The cleaned substrates were then placed in an oven to remove residual solvents. For perovskite solar cells (PSCs) with $TiO_2$ as the electron-transport layer (ETL), a compact $TiO_2$ layer was fabricated on the FTO glass substrate by spin-coating 0.15 M titanium di-isopropoxide (75 wt% in isopropanol, Sigma-Aldrich, USA) in 1-butanol solvent at 1,000 rpm for 10 s and then at 2,000 rpm for 40 s. After the spin-coating process, the compact $TiO_2$ layer was annealed twice at 125 °C for 5 min and then calcined at 500 °C for 1 h. The thickness of the prepared compact $TiO_2$ layer was ~40 nm. For the PSCs with $TiO_2/C_{60}$ as the ETL, a $TiO_2$ layer was fabricated on the cleaned FTO glass substrate using the above procedure, and a thin $C_{60}$ layer (20 nm) was deposited on the $TiO_2$ layer using a vacuum thermal evaporator. For the PSCs with $C_{60}$ as the ETL, a $C_{60}$ layer (40 nm) was deposited on the cleaned FTO glass substrate using the vacuum thermal evaporator. For cross-sectional analysis of the PSCs via the EBIC and KPFM measurements, a relatively thick $TiO_2$ layer (~100 nm) was prepared using 0.3 M titanium di-isopropoxide for the $TiO_2$-based devices, and $C_{60}$ layers with the thickness of 40 and 80 nm were prepared for the $TiO_2/C_{60}$-based and $C_{60}$-based devices, respectively.

A precursor solution of methylammonium lead iodide ($MAPbI_3$) was prepared by adding 461 mg of $PbI_2$ (Alfa Aesar, USA) and 159 mg of methylammonium iodide (MAI; Xian Polymer Light Technology, China) in a mixed solvent comprising 78 mg of dimethyl sulfoxide (DMSO; Sigma-Aldrich, USA) and 0.55 mL of *N,N*-dimethylformamide (DMF; Sigma-Aldrich, USA). The solution was spin-coated on the ETL layer at 4,000 rpm for 20 s, and 0.5 mL of diethyl ether was poured on the film 8 s after spin-coating began. A transparent adduct film was produced, and it was annealed at 100 °C for 20 min. In order to prepare a solution for the hole-transport layer (HTL), 72.3 mg of 2,2′,7,7′-tetrakis(*N,N*′-di-*p*-methoxyphenylamine)-9,9'-spirobifluorene (spiro-MeOTAD; Merck KGaA, Germany) was first dissolved in 1 mL of chlorobenzene (Sigma-Aldrich, USA). Next, 28.8 μL of 4-tert-butyl pyridine and 17.5 μL of lithium bis(trifluoromethanesulfonyl)imide from a stock solution (520 mg of lithium bisimide in 1 mL of acetonitrile, 99.8% purity, Sigma-Aldrich, USA) were added to the mixture. The HTL was prepared on the MAPbI$_3$ film by spin-coating the prepared solution at 2,000 rpm for 30 s. A gold layer with a thickness of 50 nm was deposited as a counter electrode on the HTL by using the vacuum thermal evaporator. All spin-coating processes were carried out in a dry room (<15% relative humidity, at room temperature).

*Characterization*

The current–voltage characteristics of the prepared PSCs were measured using a solar simulator (Sol3A, Oriel, USA), a source meter (2400, Keithley, USA) under AM 1.5G at 100 mW cm$^{-2}$, and a mask with an active area of 0.0729 cm$^2$ at room temperature inside a glove box. The light intensity was calibrated with a Si reference cell (Rc-1000-TC-KG5-N, VLSI Standards, USA). All EBIC measurements were performed with a field-emission scanning electron microscope (FE-SEM; Inspect F, FEI Corp., USA) using an EBIC system (DISS 5, Point Electronic GmbH, Germany). For cross-sectional EBIC imaging, an accelerating voltage of 2 kV and a working distance of 11–12 mm were used, which produced a beam current in the range of 8–25 pA. The PSCs were mechanically cleaved for the cross-sectional EBIC analysis. The cleaved surface of each device was polished with a focused ion beam (FIB) system (Quanta 3D FEG, FEI Corp., USA) for the KPFM analysis. The cross-sectional potential of the PSCs was examined with a KPFM system (Park NX10, Park Systems), using two types of Si tip coated with Cr–Au (NSC 36, Mikromasch, Germany), which had resonance frequencies of 65 and 90 kHz and spring constants of 0.6 and 1 N m$^{-1}$, at a scanning rate of 0.3 Hz and a sample bias of 3 V at 17 kHz. A white light-emitting diode (LED) with a luminous flux of 160 lm (DML 802, Makita, Japan) was used as the light source. The cross-sectional images of the PSCs were obtained using a high-resolution SEM with a FIB system (AURIGA, Carl Zeiss, Germany, or Helios 650, FEI, USA). In order to measure the accumulated charges in the PSCs, charge extraction was conducted, and Fig. S4 shows the detailed procedure. We carried out the charge extraction with different delay times ranging from 0.3 ms to 180 s, and a cluster of white LEDs with a power density of 100 mW cm$^{-2}$ was used as the light source. Impedance spectroscopy (IS) measurements were performed at the frequency range of 0.1 Hz to 1.0 MHz under ambient conditions; a cluster of white LEDs with a power density of 100 mW cm$^{-2}$ was used as the light source. We applied the bias voltage ranging from 0.0 to 1.0 V in steps of 0.2 V. The impedance spectra were fitted using ZView software (Scribner Associates, USA). The charge extraction and IS analysis were carried out using an electrochemical workstation (Autolab 320N, Metrohm, Switzerland) with an Autolab LED Driver Kit (Metrohm, Switzerland).

## Acknowledgements

This work was supported by the Global Frontier R&D Program of the Center for Multiscale Energy Systems funded by the National Research Foundation under the Ministry of Education, Science and Technology, Korea (2012M3A6A7054855).

# References

1. Q. Jiang, Y. Zhao, X. Zhang, X. Yang, Y. Chen, Z. Chu, Q. Ye, X. Li, Z. Yin and J. You, *Nature Photonics*, 2019, DOI: 10.1038/s41566-019-0398-2.
2. E. H. Jung, N. J. Jeon, E. Y. Park, C. S. Moon, T. J. Shin, T.-Y. Yang, J. H. Noh and J. Seo, *Nature*, 2019, **567**, 511-515.
3. M. M. Lee, J. Teuscher, T. Miyasaka, T. N. Murakami and H. J. Snaith, *Science*, 2012, **338**, 643.
4. W.-J. Yin, T. Shi and Y. Yan, *Advanced Materials*, 2014, **26**, 4653-4658.
5. P. Cui, D. Wei, J. Ji, H. Huang, E. Jia, S. Dou, T. Wang, W. Wang and M. Li, *Nature Energy*, 2019, **4**, 150-159.
6. B. Dänekamp, C. Müller, M. Sendner, P. P. Boix, M. Sessolo, R. Lovrincic and H. J. Bolink, *The Journal of Physical Chemistry Letters*, 2018, **9**, 2770-2775.
7. Q. Wang, Y. Shao, H. Xie, L. Lyu, X. Liu, Y. Gao and J. Huang, *Applied Physics Letters*, 2014, **105**, 163508.
8. Q. Ou, Y. Zhang, Z. Wang, J. A. Yuwono, R. Wang, Z. Dai, W. Li, C. Zheng, Z.-Q. Xu, X. Qi, S. Duhm, N. V. Medhekar, H. Zhang and Q. Bao, *Advanced Materials*, 2018, **30**, 1705792.
9. S. Olthof and K. Meerholz, *Scientific Reports*, 2017, **7**, 40267.
10. A. J. Pearson, G. E. Eperon, P. E. Hopkinson, S. N. Habisreutinger, J. T.-W. Wang, H. J. Snaith and N. C. Greenham, *Advanced Energy Materials*, 2016, **6**.
11. N. Aristidou, C. Eames, I. Sanchez-Molina, X. Bu, J. Kosco, M. S. Islam and S. A. Haque, *Nature Communications*, 2017, **8**, 15218.
12. J. S. Yun, J. Kim, T. Young, R. J. Patterson, D. Kim, J. Seidel, S. Lim, M. A. Green, S. Huang and A. Ho-Baillie, *Advanced Functional Materials*, 2018, **28**, 1705363.
13. D. Bryant, N. Aristidou, S. Pont, I. Sanchez-Molina, T. Chotchunangatchaval, S. Wheeler, J. R. Durrant and S. A. Haque, *Energy & Environmental Science*, 2016, **9**, 1655-1660.
14. N. Ahn, K. Kwak, M. S. Jang, H. Yoon, B. Y. Lee, J.-K. Lee, P. V. Pikhitsa, J. Byun and M. Choi, *Nature Communications*, 2016, **7**, 13422.
15. M.-c. Kim, N. Ahn, E. Lim, Y. U. Jin, Peter V. Pikhitsa, J. Heo, S. K. Kim, H. S. Jung

and M. Choi, *Journal of Materials Chemistry A*, 2019, **7**, 12075-12085.

16. K. Kwak, E. Lim, N. Ahn, J. Heo, K. Bang, S. K. Kim and M. Choi, *Nanoscale*, 2019, **11**, 11369-11378.
17. T. Zhang, X. Meng, Y. Bai, S. Xiao, C. Hu, Y. Yang, H. Chen and S. Yang, *Journal of Materials Chemistry A*, 2017, **5**, 1103-1111.
18. Y. Lin, B. Chen, Y. Fang, J. Zhao, C. Bao, Z. Yu, Y. Deng, P. N. Rudd, Y. Yan, Y. Yuan and J. Huang, *Nature Communications*, 2018, **9**, 4981.
19. Y. Han, S. Meyer, Y. Dkhissi, K. Weber, J. M. Pringle, U. Bach, L. Spiccia and Y.-B. Cheng, *Journal of Materials Chemistry A*, 2015, **3**, 8139-8147.
20. F. Matteocci, L. Cinà, E. Lamanna, S. Cacovich, G. Divitini, P. A. Midgley, C. Ducati and A. Di Carlo, *Nano Energy*, 2016, **30**, 162-172.
21. E. Edri, S. Kirmayer, A. Henning, S. Mukhopadhyay, K. Gartsman, Y. Rosenwaks, G. Hodes and D. Cahen, *Nano Letters*, 2014, **14**, 1000-1004.
22. N. Kedem, M. Kulbak, T. M. Brenner, G. Hodes and D. Cahen, *Physical Chemistry Chemical Physics*, 2017, **19**, 5753-5762.
23. N. Kedem, T. M. Brenner, M. Kulbak, N. Schaefer, S. Levcenko, I. Levine, D. Abou-Ras, G. Hodes and D. Cahen, *The Journal of Physical Chemistry Letters*, 2015, **6**, 2469-2476.
24. W. S. Lau, D. S. H. Chan, J. C. H. Phang, K. W. Chow, K. S. Pey, Y. P. Lim, V. Sane and B. Cronquist, *Journal of Applied Physics*, 1995, **77**, 739-746.
25. O. Breitenstein, J. Bauer, M. Kittler, T. Arguirov and W. Seifert, *Scanning*, 2008, **30**, 331-338.
26. M. Kulbak, I. Levine, E. Barak-Kulbak, S. Gupta, A. Zohar, I. Balberg, G. Hodes and D. Cahen, *Advanced Energy Materials*, 2018, **8**, 1800398.
27. O. Hentz, P. Rekemeyer and S. Gradečak, *Advanced Energy Materials*, 2018, **8**, 1701378.
28. V. W. Bergmann, S. A. L. Weber, F. Javier Ramos, M. K. Nazeeruddin, M. Grätzel, D. Li, A. L. Domanski, I. Lieberwirth, S. Ahmad and R. Berger, *Nature Communications*, 2014, **5**, 5001.
29. S. A. L. Weber, I. M. Hermes, S.-H. Turren-Cruz, C. Gort, V. W. Bergmann, L. Gilson, A. Hagfeldt, M. Graetzel, W. Tress and R. Berger, *Energy & Environmental Science*,

2018, **11**, 2404-2413.

30. V. W. Bergmann, Y. Guo, H. Tanaka, I. M. Hermes, D. Li, A. Klasen, S. A. Bretschneider, E. Nakamura, R. Berger and S. A. L. Weber, *ACS Applied Materials & Interfaces*, 2016, **8**, 19402-19409.

31. Q. Chen, L. Mao, Y. Li, T. Kong, N. Wu, C. Ma, S. Bai, Y. Jin, D. Wu, W. Lu, B. Wang and L. Chen, *Nature Communications*, 2015, **6**, 7745.

32. C.-W. Chu, V. Shrotriya, G. Li and Y. Yang, *Applied Physics Letters*, 2006, **88**, 153504.

33. Y. Li, Y. Zhao, Q. Chen, Y. M. Yang, Y. Liu, Z. Hong, Z. Liu, Y. T. Hsieh, L. Meng, Y. Li and Y. Yang, *J Am Chem Soc*, 2015, **137**, 15540-15547.

34. J. Bisquert, F. Fabregat-Santiago, I. Mora-Seró, G. Garcia-Belmonte and S. Giménez, *The Journal of Physical Chemistry C*, 2009, **113**, 17278-17290.

35. R. K. Misra, S. Aharon, M. Layani, S. Magdassi and L. Etgar, *Journal of Materials Chemistry A*, 2016, **4**, 14423-14429.

36. F. Giordano, A. Abate, J. P. Correa Baena, M. Saliba, T. Matsui, S. H. Im, S. M. Zakeeruddin, M. K. Nazeeruddin, A. Hagfeldt and M. Graetzel, *Nature Communications*, 2016, **7**, 10379.

37. H.-S. Kim, I.-H. Jang, N. Ahn, M. Choi, A. Guerrero, J. Bisquert and N.-G. Park, *The Journal of Physical Chemistry Letters*, 2015, **6**, 4633-4639.

38. I. Zarazua, J. Bisquert and G. Garcia-Belmonte, *The Journal of Physical Chemistry Letters*, 2016, **7**, 525-528.

39. I. Zarazua, G. Han, P. P. Boix, S. Mhaisalkar, F. Fabregat-Santiago, I. Mora-Sero, J. Bisquert and G. Garcia-Belmonte, *J Phys Chem Lett*, 2016, **7**, 5105-5113.

40. E. Bi, H. Chen, F. Xie, Y. Wu, W. Chen, Y. Su, A. Islam, M. Grätzel, X. Yang and L. Han, *Nature Communications*, 2017, **8**, 15330.

41. Z. Wang, D. P. McMeekin, N. Sakai, S. van Reenen, K. Wojciechowski, J. B. Patel, M. B. Johnston and H. J. Snaith, *Advanced Materials*, 2017, **29**, 1604186.

42. Y. C. Zhao, W. K. Zhou, X. Zhou, K. H. Liu, D. P. Yu and Q. Zhao, *Light Sci Appl*, 2017, **6**, e16243.

43. H.-S. Kim, I. Mora-Sero, V. Gonzalez-Pedro, F. Fabregat-Santiago, E. J. Juarez-Perez, N.-G. Park and J. Bisquert, *Nature Communications*, 2013, **4**, 2242.

44. Y. Deng, Z. Xiao and J. Huang, *Advanced Energy Materials*, 2015, **5**, 1500721.

45. K. Domanski, B. Roose, T. Matsui, M. Saliba, S.-H. Turren-Cruz, J.-P. Correa-Baena, C. R. Carmona, G. Richardson, J. M. Foster, F. De Angelis, J. M. Ball, A. Petrozza, N. Mine, M. K. Nazeeruddin, W. Tress, M. Grätzel, U. Steiner, A. Hagfeldt and A. Abate, *Energy & Environmental Science*, 2017, **10**, 604-613.
46. W. Nie, J.-C. Blancon, A. J. Neukirch, K. Appavoo, H. Tsai, M. Chhowalla, M. A. Alam, M. Y. Sfeir, C. Katan, J. Even, S. Tretiak, J. J. Crochet, G. Gupta and A. D. Mohite, *Nature Communications*, 2016, **7**, 11574.
47. G. Ahmad, S. Mandal, A. K. Barua, T. K. Bhattacharya and J. N. Roy, *IEEE Journal of Photovoltaics*, 2017, **7**, 414-420.
48. T. Söderström, F. J. Haug, V. Terrazzoni-Daudrix and C. Ballif, *Journal of Applied Physics*, 2008, **103**, 114509.

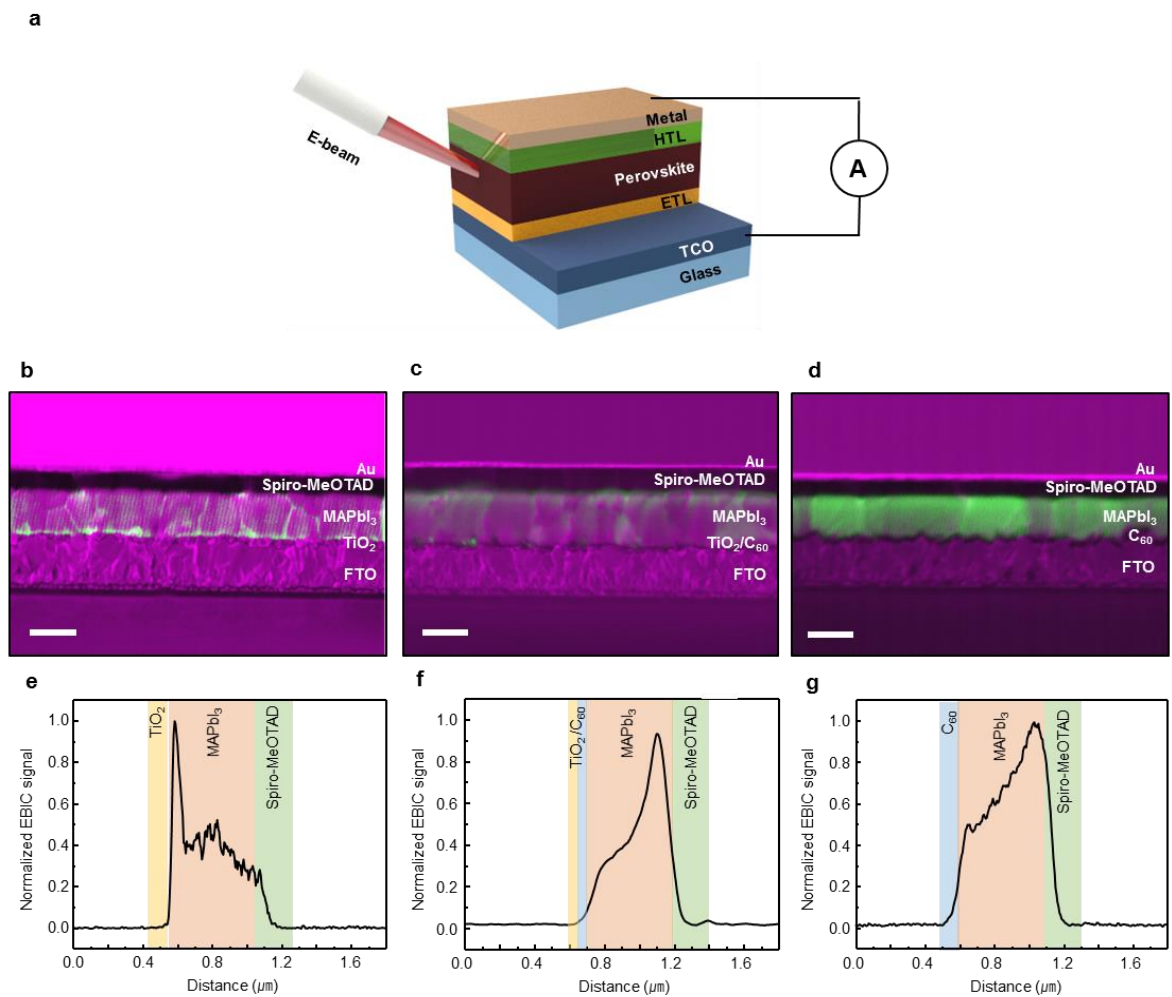

**Fig. 1.** (a) Schematic illustration of EBIC measurement on a cross-section of a perovskite solar cell. Cross-sectional EBIC image overlapped with SEM image of (b) $TiO_2$-based, (c) $TiO_2/C_{60}$-based, and (d) $C_{60}$-based devices. The EBIC signal is overlaid in a bright green color on the SEM image. Line profiles of EBIC signal perpendicular to layers of the (e) $TiO_2$-based, (f) $TiO_2/C_{60}$-based, and (g) $C_{60}$-based devices. Scale bars: 500 nm.

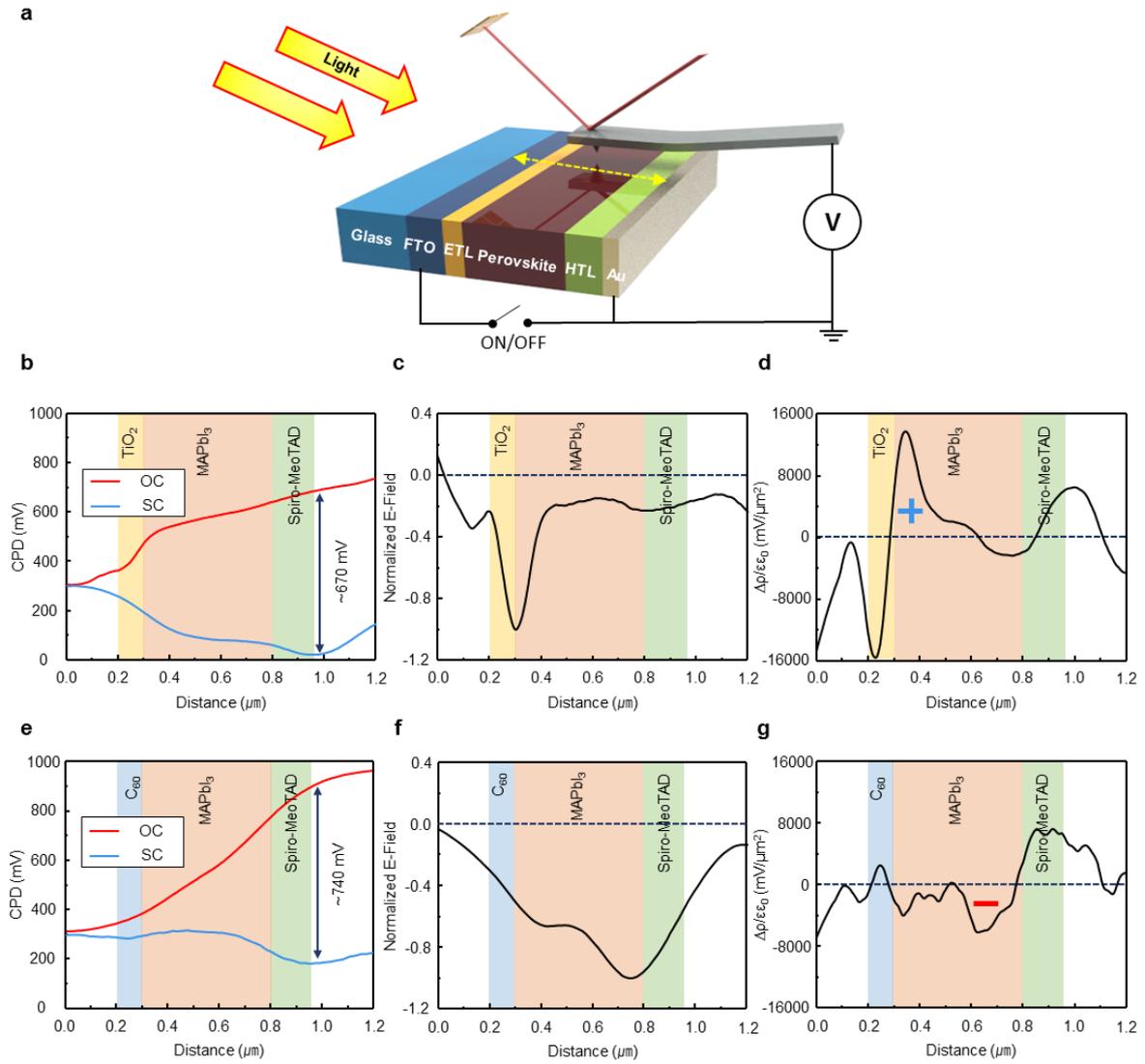

**Fig. 2.** (a) Schematic illustration of cross-sectional KPFM measurement under operating conditions. CPD distribution under illumination under open-circuit (OC) condition (red line) and under short-circuit (SC) condition (blue line) of (b) $TiO_2$-based and (e) $C_{60}$-based devices. The arrows in (b) and (e) mark the built-up open-circuit voltage, $V_{OC}$. Normalized E-field distribution under illumination and open-circuit condition of the (c) $TiO_2$-based and (f) $C_{60}$-based devices. Calculated charge density profile obtained from photo-potential difference between the open-circuit and short-circuit voltages ($V_{OC} - V_{SC}$) of (d) $TiO_2$-based and (g) $C_{60}$-based devices.

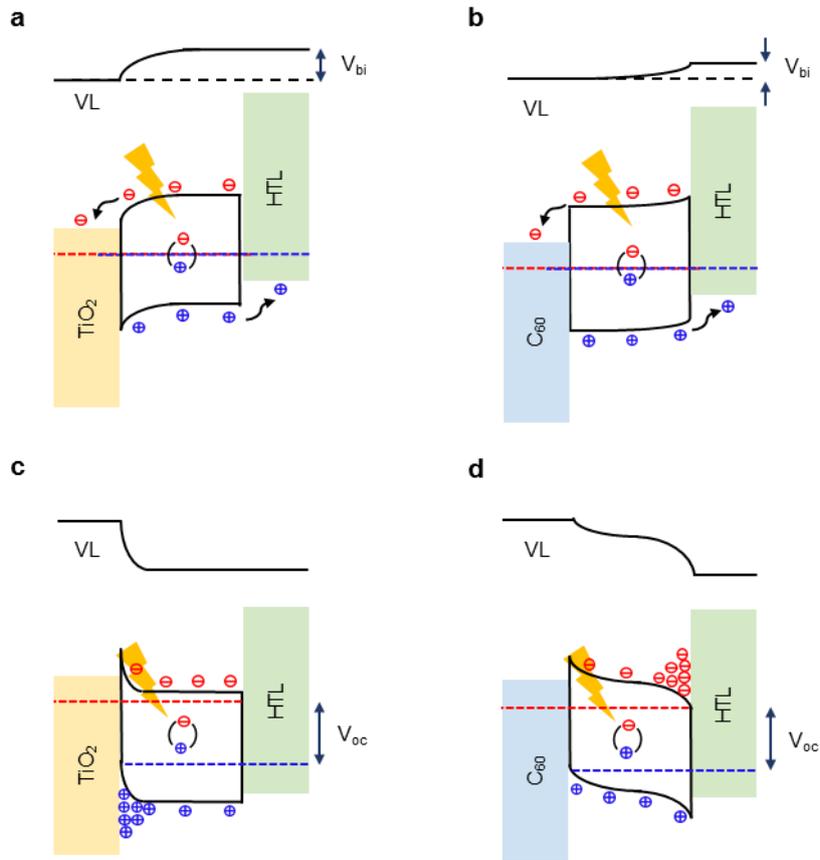

**Fig. 3.** Band diagrams of TiO$_2$-based device under light illumination in (a) short-circuit state and (c) open-circuit state. Band diagrams of the C$_{60}$-based device under light illumination with both (b) short-circuit state and (d) open-circuit state. Here, the red dotted line and blue dotted line represent the quasi-electron Fermi level and quasi-hole Fermi level, respectively. VL: vacuum level; HTL, hole-transport layer; $V_{BI}$, built-in potential; $V_{OC}$, open-circuit voltage

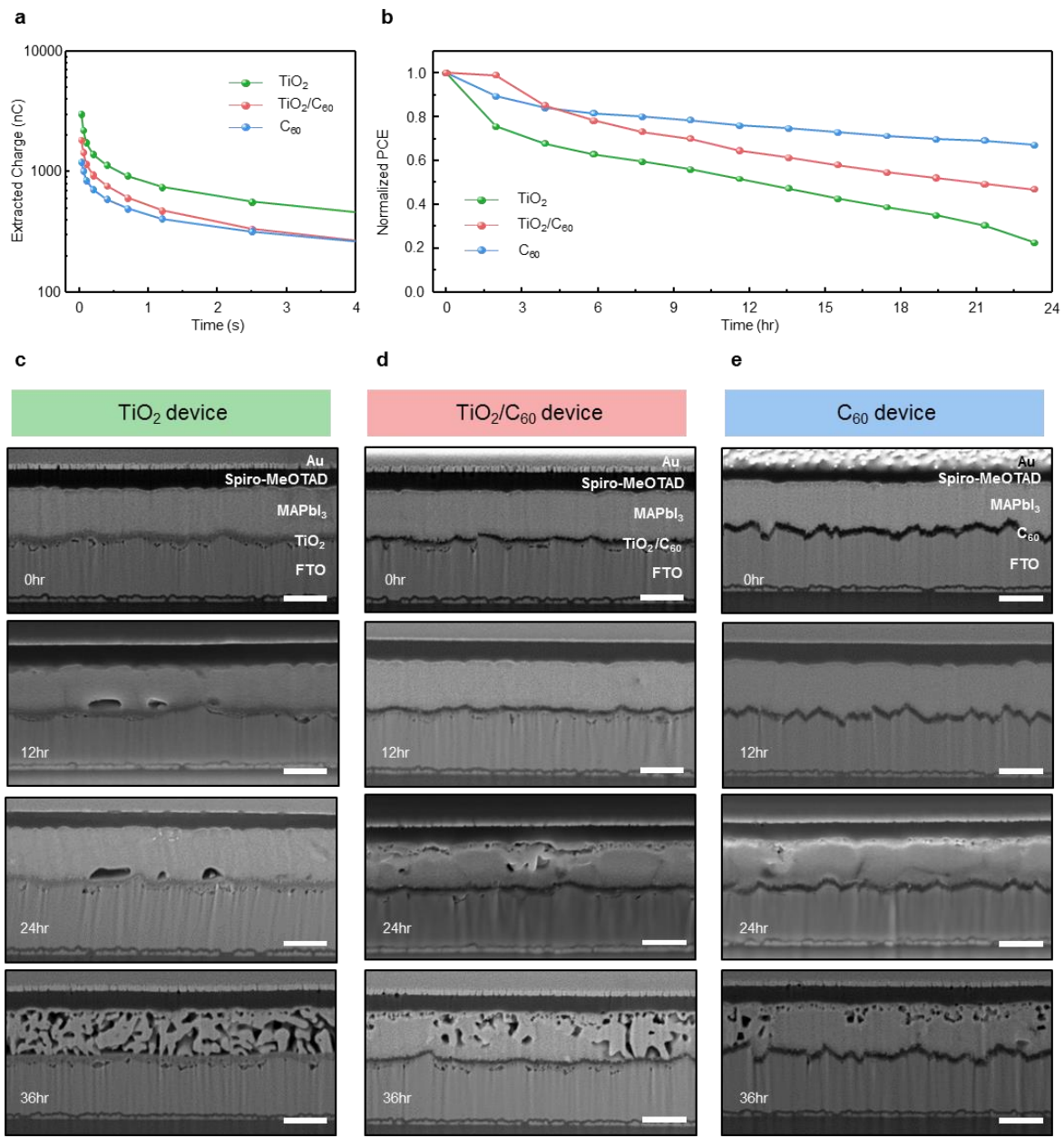

**Fig. 4.** (a) Amount of extracted charges as a function of switching delay time (up to 4 s) for PSCs with different types of electron-transport layer (ETL). (b) Time evolution of normalized power-conversion efficiencies (PCEs) of non-encapsulated devices under 1 sun illumination in air. Time evolution of cross-sectional SEM images of (c) $TiO_2$-based, (d) $TiO_2/C_{60}$-based, and (e) $C_{60}$-based devices during device degradation. Scale bars: 500 nm

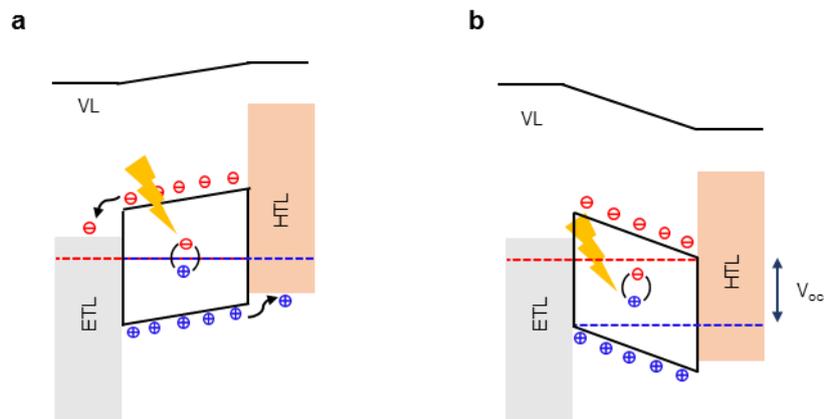

**Fig. 5.** Schematic illustrations of band diagrams of device with ideal p–i–n junction type under light illumination in (a) short-circuit state and (b) open-circuit state. VL: vacuum level; ETL, electron-transport layer; HTL, hole-transport layer; $V_{OC}$, open-circuit voltage

# Charge transport layer dependent electronic band bending in perovskite solar cells and its correlation to device degradation

Junseop Byeon[1,2,†], Jutae Kim[1,†], Ji-Young Kim[3], **Gunhee Lee**[1]**, Kijoon Bang**[1], Namyoung Ahn*[1], and Mansoo Choi*[1,2]

[1]Global Frontier Center for Multiscale Energy Systems, Seoul National University, Seoul 08826, Republic of Korea.
[2]Department of Mechanical and Aerospace Engineering, Seoul National University, Seoul 08826, Republic of Korea.
[3]Advanced Analysis Center, Korea Institute of Science and Technology (KIST), Hwarangno 14-gil 5, Seongbuk-gu, Seoul 02792, Republic of Korea.

* To whom correspondence should be addressed.
E-mail: Mansoo Choi (mchoi@snu.ac.kr), Namyoung Ahn(nyny92@snu.ac.kr)

† These authors contributed equally to this work.

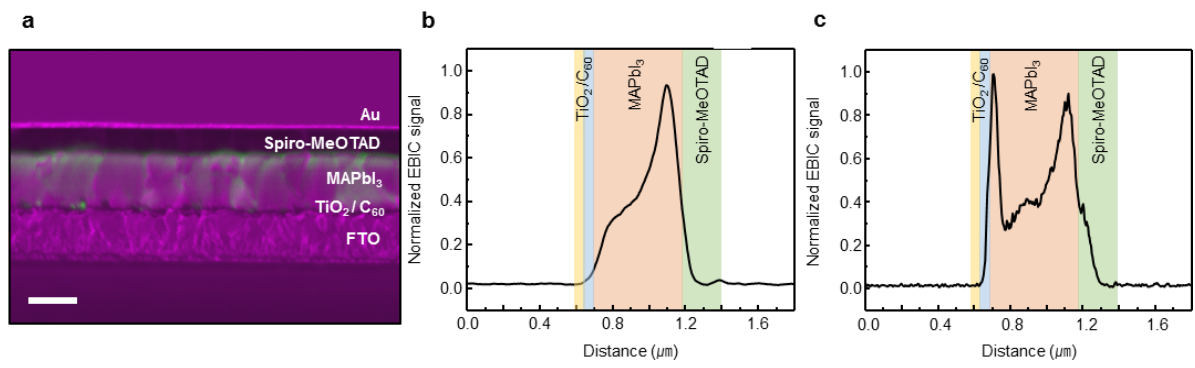

**Fig. S1.** (a) Cross sectional EBIC image overlapped with SEM image of the $TiO_2/C_{60}$-based device. (b) Line profiles of EBIC signal perpendicular to layers of the regions without (b) and with a local pinhole in the $C_{60}$ layer in the device. Scale bar: 500nm.

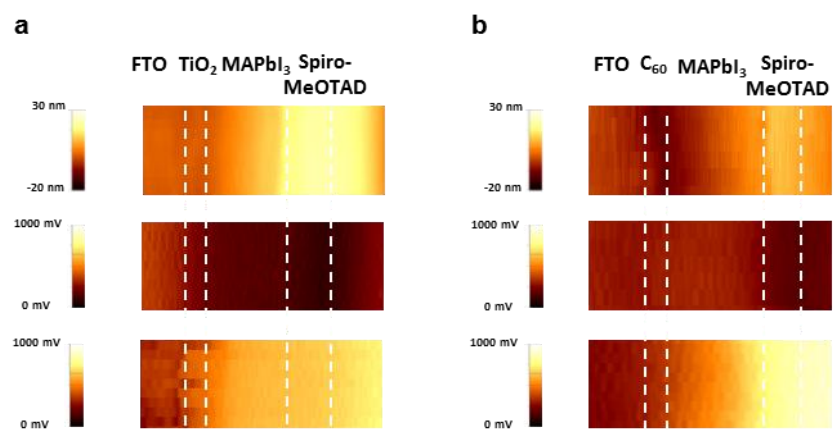

**Fig. S2.** (a) A topographic image (Top), CPD distribution images under short-circuit (middle), and open circuit (bottom) conditions of the $TiO_2$-based device. (b) A topographic image (Top), CPD distribution images under short-circuit (middle), and open-circuit (bottom) conditions of the $C_{60}$-based device. RMS of roughness in the topographic images of the $TiO_2$-based and $C_{60}$-based devices are 9.32 nm, 5.55 nm, respectively.

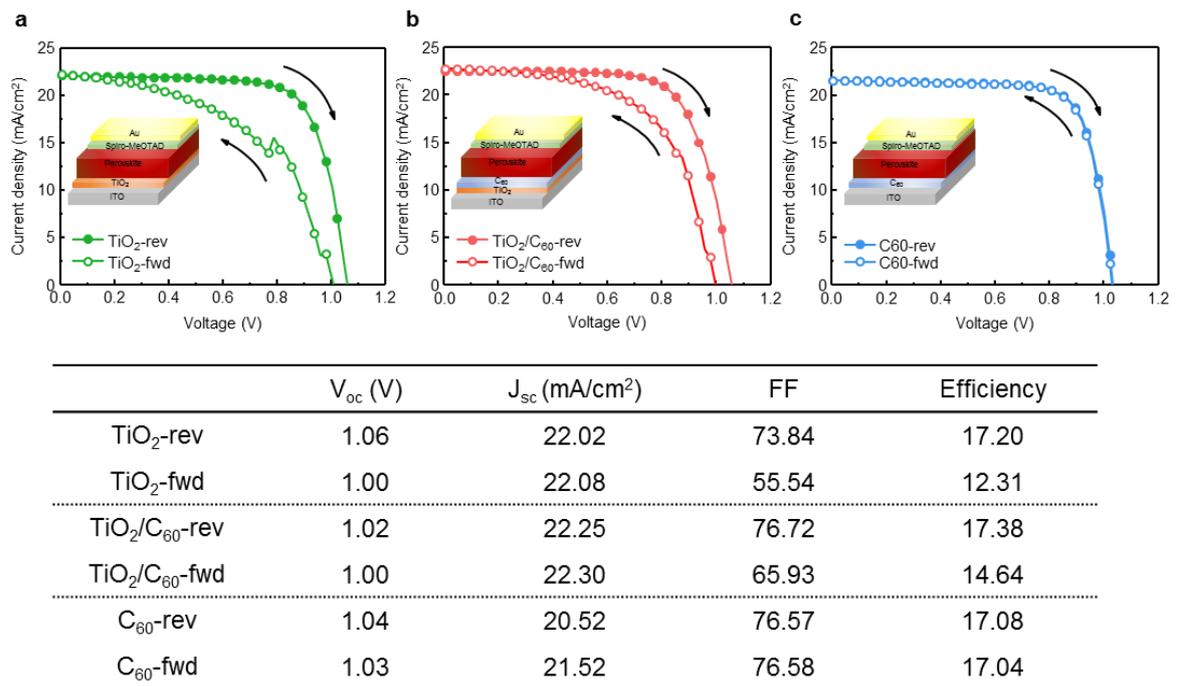

|  | $V_{oc}$ (V) | $J_{sc}$ (mA/cm$^2$) | FF | Efficiency |
|---|---|---|---|---|
| TiO$_2$-rev | 1.06 | 22.02 | 73.84 | 17.20 |
| TiO$_2$-fwd | 1.00 | 22.08 | 55.54 | 12.31 |
| TiO$_2$/C$_{60}$-rev | 1.02 | 22.25 | 76.72 | 17.38 |
| TiO$_2$/C$_{60}$-fwd | 1.00 | 22.30 | 65.93 | 14.64 |
| C$_{60}$-rev | 1.04 | 20.52 | 76.57 | 17.08 |
| C$_{60}$-fwd | 1.03 | 21.52 | 76.58 | 17.04 |

**Fig. S3.** J-V characteristic of the (a) TiO$_2$-based, (b) TiO$_2$/C$_{60}$-based, (c) C$_{60}$-based devices measured in reverse (full circle) and forward (hollow circle) scan.

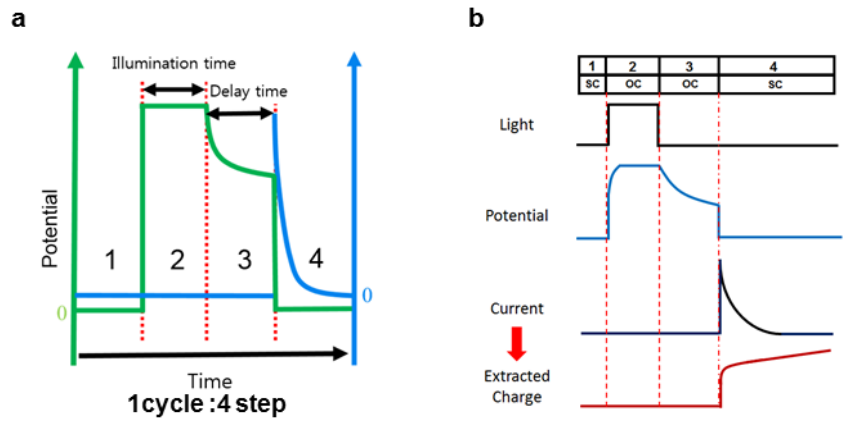

**Fig. S4.** The sequence of the charge extraction measurement: 1) light off & short-circuit (SC) step in which charge remained in the device is extracted from the previous cycle; 2) light on & pen-circuit (OC) step in which photocurrent is generated and carriers are recombined; 3) light off & OC step in which charge generation is stopped and accumulated charge inside the device is relaxed and recombined under certain delay time; 4) light off & SC step in which remnant accumulated charge is extracted. (a) An illustration of the potential and current values as a function of the steps, and (b) values of the extracted charge by integration of the current profile.

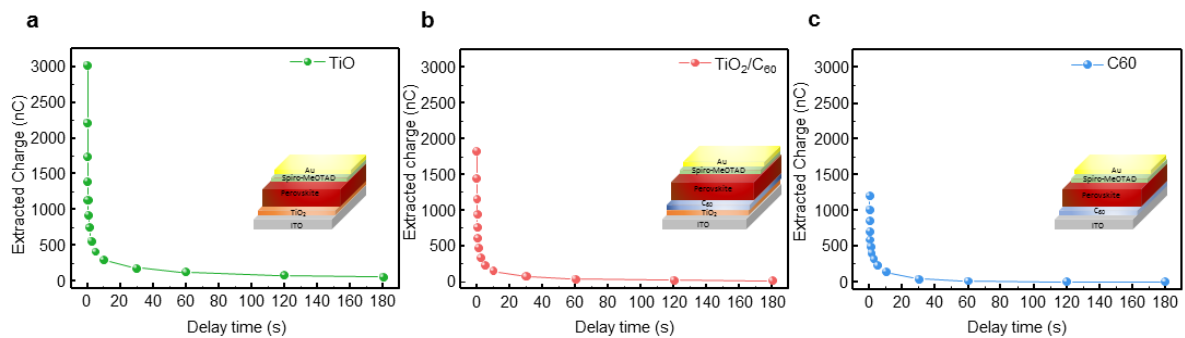

| Q (nC) | 0.03 s | 0.06 s | 0.1 s | 0.4 s | 1.2 s | 5 s | 10 s | 30 s | 60 s | 180 s |
|---|---|---|---|---|---|---|---|---|---|---|
| $TiO_2$ | 3025.52 | 2213.63 | 1738.15 | 1134.74 | 747.73 | 410.59 | 296.04 | 183.12 | 124.87 | 56.33 |
| $TiO_2/C_{60}$ | 1833.67 | 1448.08 | 1154.97 | 762.99 | 476.79 | 229.69 | 152.36 | 73.60 | 41.13 | 13.80 |
| $C_{60}$ | 1207.13 | 1012.23 | 849.92 | 591.36 | 406.43 | 229.33 | 143.29 | 40.44 | 12.18 | 2.04 |

**Fig. S5.** The amount of extracted charges as a function of the delay time up to 180 s of (a) the $TiO_2$-based, (b) $TiO_2/C_{60}$-based, (c) $C_{60}$-based devices.

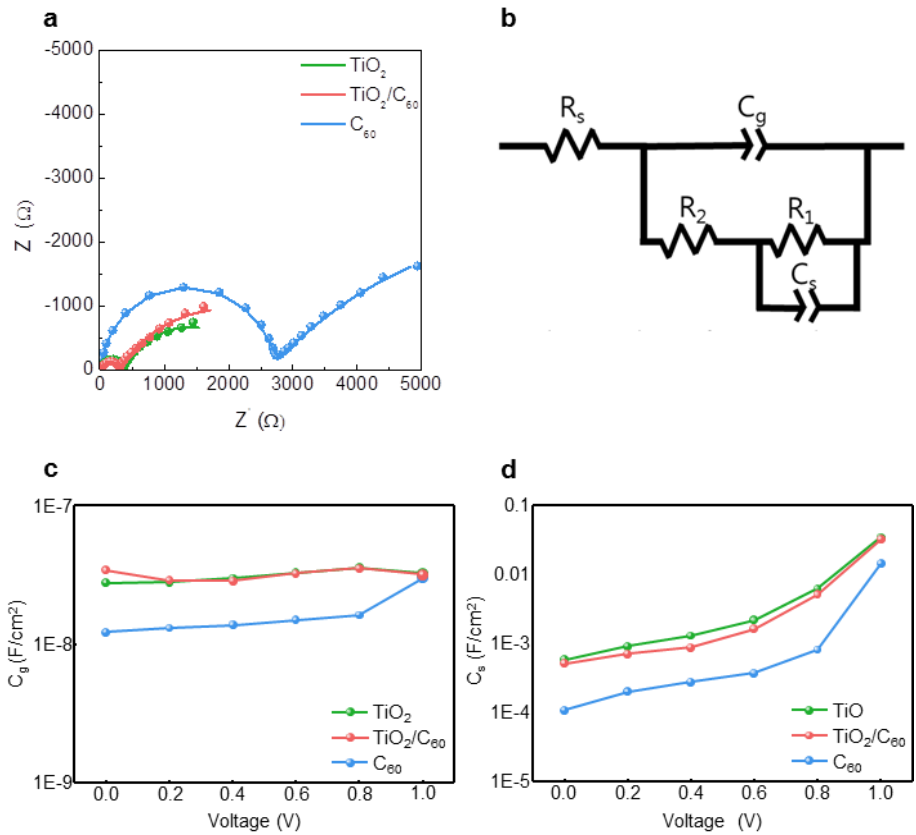

**Fig. S6.** (a) Nyquist plots (Z"-Z') of the PSCs with different ETLs. (b) An equivalent circuit employed in this study and (c) the values of the geometrical capacitance and (d) surface accumulation capacitance depending on the applied voltage. EIS measurements are performed under a LED light source and ambient conditions. Here, $R_s$ is ohmic contribution of contacts and wires; $C_g$ is geometrical capacitance, which is dielectric response of the perovskite layer; $C_s$ is capacitance of surface charge accumulation at the interfaces; and $R_1$ and $R_2$ are the resistance of recombination current flux.